\documentclass[twocolumn]{aastex631}

\shorttitle{Void replenishment: how voids accrete matter}
\shortauthors{Vallés-Pérez, Quilis \& Planelles}

\defcitealias{Ricciardelli_2013}{RQP13}
\newcommand{\vb}[1]{\mathbf{#1}}

\begin{document}

\title{Void replenishment: how voids accrete matter over cosmic history\footnote{Released on \today.}}

\author[0000-0003-2656-5985]{David Vall{\'e}s-P{\'e}rez}
\affiliation{Departament d'Astronomia i Astrofísica, Universitat de València. Burjassot (València), 46100}

\author[0000-0002-2852-5031]{Vicent Quilis}
\affiliation{Departament d'Astronomia i Astrofísica, Universitat de València. Burjassot (València), 46100}
\affiliation{Observatori Astronòmic, Universitat de València. Paterna (València), 46980}

\author[0000-0002-0105-4815]{Susana Planelles}
\affiliation{Departament d'Astronomia i Astrofísica, Universitat de València. Burjassot (València), 46100}
\affiliation{Observatori Astronòmic, Universitat de València. Paterna (València), 46980}

\email{E-mail addresses: david.valles-perez@uv.es, vicent.quilis@uv.es, susana.planelles@uv.es}

\begin{abstract}

Cosmic voids are underdense regions filling up most of the volume in the Universe. They are expected  to emerge in regions comprising negative initial density fluctuations, and subsequently expand as the matter around them collapses and forms walls, filaments and clusters. We report results from the analysis of a cosmological simulation specially designed to accurately describe low-density regions, such as cosmic  voids. Contrary to the common expectation, we find that voids also experience significant mass inflows over cosmic history. On average, 10\% of the mass of voids in the sample at  $z\sim 0$ is accreted from overdense regions, reaching values beyond 35\% for a significant fraction of voids. More than half of the mass entering the voids lingers on periods of time $\sim 10\, \mathrm{Gyr}$ well inside them, reaching inner radii. This would imply that part of the gas lying inside voids at a given time proceeds from overdense regions (e.g., clusters or filaments), where it could have been pre-processed, thus challenging the scenario of galaxy formation in voids, and dissenting from the idea of being pristine environments.

\end{abstract}

\keywords{Large-scale structure of the Universe (902) --- Voids (1779) --- Accretion (14) --- Galaxy environments (2029) --- Computational astronomy (293)}

%%%%%%%%%%%%%
\section{Introduction} 
\label{sec:intro}
%%%%%%%%%%%%%

Cosmic voids are underdense regions filling up most of the volume in the Universe \citep{Zeldovich_1982}. According to the accepted paradigm of cosmological structure formation, they emerge in regions comprising negative initial density fluctuations \citep{Sheth_2004}, and subsequently expand as the matter around them collapses and forms walls, filaments and clusters (see \citealp{vandeWeygaert_2011_review} and \citealp{vandeWeygaert_2016_review} for recent, general reviews). This leads to coherent outflows \citep{vandeWeygaert_1993, Padilla_2005, Ceccarelli_2006, Patiri_2012}, making them a pristine environment with notable applications for cosmology \citep{Dekel_1994, Park_2007, Lavaux_2010, Lavaux_2012, Bos_2012, Pisani_2019} and galaxy formation \citep{Hahn_2007, vandeWeygaert_2011_review, Kreckel_2011, Ricciardelli_2014b}.

The dynamics of cosmic voids are dominated by their expansion and consequent depletion of gas and dark matter (DM), as revealed by the coherent outflows found both in simulations \citep{vandeWeygaert_1993, Padilla_2005, Ceccarelli_2006} and observations \citep{Bothun_1992, Patiri_2012, Paz_2013}, and also expected from analytical models of isolated voids \citep{Sheth_2004,Bertschinger_1985,Baushev_2021}. However, in a fully cosmological environment, it should be in principle possible to expect coherent streams of matter --gas and DM-- to unbind from dense structures and end up penetrating inside low-density regions. As a matter of fact, a handful of scenarios for unbinding mass do exist, such as galaxy cluster mergers \citep{Behroozi_2013} or strong shocks which can extend up to a few virial radii (e.g., \citealp{Zhang_2020}).

In this Letter, we explore this scenario with a $\Lambda$ cold dark matter ($\Lambda$CDM) cosmological simulation of a large volume domain, especially designed to describe matter in and around voids. The rest of the Letter is organised as follows. In Sec. \ref{sec:methods}, we describe the simulation and the void finding algorithm. In Sec. \ref{sec:results}, we present our results regarding the existence of mass inflows through voids' boundaries. Finally, we summarize the implications of these results in Sec. \ref{sec:conclusion}.

%%%%%%%%%%%%%
\section{Methods} 
\label{sec:methods}
%%%%%%%%%%%%%

%%%%%%%%%%%%%%
\subsection{The simulation}
\label{s:methods.simulation}
%%%%%%%%%%%%%%

The results reported in this paper proceed from a cosmological simulation of a periodic domain, $100 \, h^{-1} \, \mathrm{Mpc}$ along each direction, produced with \texttt{MASCLET} \citep{Quilis_2004}, an Eulerian, adaptive mesh refinement (AMR) hydrodynamics coupled to particle-mesh $N$-Body code. The Eulerian hydrodynamic scheme in \texttt{MASCLET}, based on high-resolution shock-capturing techniques, is capable of providing a faithful description of the gaseous component in low-density regions, such as cosmic voids. 

Structures evolve on top of a flat $\Lambda$CDM cosmology consistent with the latest \cite{Planck_2020} results. Dark energy, matter and baryon densities are specified by $\Omega_\Lambda = 0.69$, $\Omega_m = 0.31$, $\Omega_b = 0.048$, relative to the critical density $\rho_c = \frac{3 H_0^2}{8\pi G}$. The Hubble parameter, $H_0 = 100 \, h \, \mathrm{km\, s^{-1}\, Mpc^{-1}}$, is set by $h=0.678$. The initial conditions were set up at at $z=100$, by evolving a power spectrum realization with spectral index $n_s = 0.96$ and normalization $\sigma_8 = 0.82$ using \citeauthor{Zeldovich_1970}'s (\citeyear{Zeldovich_1970}) approximation.

A low-resolution run on a grid of $128^3$ cells was performed in order to identify the regions which would evolve into cosmic voids by $z \simeq 0$. Back to the initial conditions, the seeds of voids and their surroundings were sampled with higher numerical resolution according to the procedure introduced in \citet[hereon, \citetalias{Ricciardelli_2013}]{Ricciardelli_2013}. The regions at $z=100$ comprising the DM particles which end up in zones with\footnote{$\rho_B \equiv \Omega_m \rho_c$ is the \textit{background matter density} of the Unvierse.} $\rho/\rho_B < 10$ by $z=0$ are thus mapped with a first level of mesh refinement ($\ell = 1$), with half the cell size and DM particles eight times lighter than those of the base grid, therefore with DM mass resolution $6.4 \times 10^9 M_{\odot}$ and spatial resolution $390 h^{-1} \, \mathrm{kpc}$. In order to capture the structures forming in cosmic voids, subsequent levels of refinement ($\ell \geq 2$, up to $n_\ell = 10$) are created following a pseudo-Lagrangian approach that refines cells where density has increased a factor of eight with respect to the previous, lower-resolution level. Besides gravity and hydrodynamics, the simulation includes standard cooling and heating mechanisms, and a phenomenological parametrisation of star formation \citep{Quilis_2017}.

%%%%%%%%%%%%%%
\subsection{The void finder}
\label{s:methods.void finder}
%%%%%%%%%%%%%%

We have identified the sample of cosmic voids in our simulation with a void finder based on the one presented by \citetalias{Ricciardelli_2013}, which looks for ellipsoidal voids using the total density field ($\rho_\mathrm{tot}$) and the gas velocity field ($\vb{v}$), as underdense ($\rho_\mathrm{tot}<\rho_\mathrm{B}$), peculiarly expanding ($\nabla \cdot \vb{v} > 0$) regions surrounded by steep density gradients. While the original void finder in \citetalias{Ricciardelli_2013} did not assume any prior on the voids shape, which could therefore develop highly complex, non-convex and non-simply connected shapes, such a precise definition of a void boundary is counter-productive for assessing mass fluxes in post-processing, since it limits the validity of the pseudo-Lagrangian approach (see Sec. \ref{s:methods.pseudolagrangian}).

In order to have voids with smooth surfaces, our void finder looks for voids as ellipsoidal volumes around density minima, using the same thresholds on total density, total density gradient and gas velocity divergence as \citetalias{Ricciardelli_2013}. While voids are not generally ellipsoidal, by using the same threshold values as \citetalias{Ricciardelli_2013} we ensure that our algorithm looks for the largest possible ellipsoid inside actual, complex-shaped voids, thus providing a robust, stable and conservative definition of these structures which is readily comparable with their identification in observational data \citep{Foster_2009, Patiri_2012}. The voids are found and characterised one at a time, using the $128^3$ base grid, in the steps summarised below.

\paragraph*{Protovoid finding.$\;$} A tentative center is chosen, as the most underdense, positive velocity divergence cell not yet inside an already found void. The initial protovoid is a $5^3$ cells cube around this cell. The protovoid is then grown iteratively in the directions of its six faces, one extra cell along each direction at a time, if the following conditions are met by all the `new' cells:

\begin{equation}
	\delta_\mathrm{tot} < 0, \qquad \nabla \delta_\mathrm{tot} < \left( \nabla \delta \right)^\mathrm{max}, \qquad \nabla \cdot \vb{v} > 0,
	\label{eq:voidfind.conditions.protovoid}
\end{equation}

\noindent where $\delta_\mathrm{tot} \equiv \frac{\rho_\mathrm{tot}}{\rho_B} - 1$ is the total density contrast. The threshold on the density gradient is set to $\left( \nabla \delta \right)^\mathrm{max} = 0.25 \, \mathrm{Mpc^{-1}}$, in consistency with \citetalias{Ricciardelli_2013}. Once the protovoid has been determined, the center of the void is adjusted to the center of mass-defect of the protovoid. A first approximation to the shape of the ellipsoid is got by computing and diagonalizing the inertia tensor.

\paragraph*{Growth of the ellipsoidal void.$\;$} The initial ellipsoid is subsequently grown iteratively to find the maximal ellipsoid that fits inside the actual void. This is performed by repeatedly applying the following two substeps:

\begin{enumerate}
	\item The shape of the ellipsoid is adjusted iteratively, in a similar manner to what is done in the galaxy clusters literature \citep{Zemp_2011, Valles-Perez_2020}. In particular, the new eigenvalues of the inertia tensor yield the orientation of the void, while the new eigenvectors are used to compute the new semiaxes. These semiaxes are rescaled proportionally, so as to preserve the volume of the ellipsoid. The process is iterated, adjusting the integration volume, until convergence, which is assessed by the change in semiaxes lengths. 
	\item Once the shape has been found, the ellipsoidal void is grown at constant shape, by multiplying each of its semiaxes by a factor $1+\chi$. We have fixed $\chi=0.05$, although this parameter does not have a severe impact on the resulting void population while kept small.
\end{enumerate}

This two-step iteration is repeated until either one of the stopping conditions in \citetalias{Ricciardelli_2013} (maximum density, density gradient or negative velocity diverenge) is met by a cell, or the mean slope of the total density field at the boundary exceeds the prediction of the universal density profile \citep{Ricciardelli_2013, Ricciardelli_2014}. It can be easily shown that, assuming that the spherical profiles in \citetalias{Ricciardelli_2013} can be extended to ellipsoidal shells, this condition can be applied by requiring:

\begin{equation}
	3 \left(\frac{\rho(R)}{\rho(< R)} - 1\right) > 1.37 - 0.25 z
	\label{eq:voidfind.slope_criterion}
\end{equation}

\noindent with $\rho(<R)$ the mean density inside the ellipsoid and $\rho(R)$ computed from the newly added cells in the growing step. The numerical coefficients, which are derived from the fit in \citetalias{Ricciardelli_2013}, are valid for $2.5 \geq z \geq 0$.

\paragraph*{Void sample and merger tree.$\;$} In order to produce the final sample of voids, we start from the latest code output (at $z=0$) and select an initial sample of voids taking care of the overlaps. To do so, we iterate through the voids found by the algorithm described above, from the largest and emptiest to the smallest and densest, and accept those which do not overlap more than $50\%$ with the volume occupied by previously accepted voids. 

Then, we trace this initial sample back in time by building their merger tree. To assess which is the best progenitor candidate for a void, we find the one which maximises the \textit{volume retention}\ defined as $VR \equiv \frac{N_\mathrm{A \cap B}}{\sqrt{N_\mathrm{A} N_\mathrm{B}}}$, with $N$ referring to the number of cells, and $A$, $B$ being some void and one of its parent candidates, respectively.
This approach is equivalent to the \textit{particle retention} defined by \citet{Minoguchi_2021}, but here applied to cells \citep{Sutter_2014}. We did not find strong overall variations when using other figures of merit defined on \citet{Minoguchi_2021}, although there can be variations on a small number of individual voids.

With this procedure, we are able to obtain a sample of 207 voids, with equivalent radius at $z=0$ larger than 5 Mpc (the largest of them reaching $\sim 18\, \mathrm{Mpc}$), which can be traced back, at least, to $z = 1.5$; and 179 of them are traced back down to $z=2.5$. The overall statistics (radii, elipticities and mean overdensities) of the void sample are displayed in Supplementary Figure \ref{fig:supplementary_sample}.

%%%%%%%%%%%%%%
\subsection{The pseudo-Lagrangian approach and its validity}
\label{s:methods.pseudolagrangian}
%%%%%%%%%%%%%%

To assess mass fluxes in post-processing, we take a pseudo-Lagrangian approach, by interpreting each volume element in the simulation (up to a certain refinement level) as a tracer particle, and advecting these tracer particles using the gas velocity field between each pair of code snapshots. This technique is analogue to the one applied by \citet{Valles-Perez_2020} for galaxy clusters.

In practical terms, at each code snapshot we take all (non-refined and non-overlapping) gas cells, each one at a position $\vb{x_\mathrm{n}}$ and compute their updated position with an explicit, first order step. For consistency, we have performed the same analysis with the dark matter particle distribution. Since voids experience mild dynamics with large dynamical times (associated to their low densities), this is a sensible approach. Then, in order to compute the accretion mass flux around a void between a pair of code outputs, we consider the mass of all the dark matter particles (or gas pseudo-particles) which were outside the void in the previous iteration, and inside the same volume at the latter (and viceversa for the decretion mass flux).

The previously discussed procedure will be valid as long as the timestep involved (the timespan between two consecutive outputs of the simulation) fulfils $\Delta t \lesssim \mathcal{L} / \mathcal{V}$, with $\mathcal{L}$ a characteristic scale of the surface of the void, and $\mathcal{V}$ a characteristic velocity for these cosmic flows. By choosing an ellipsoidal, instead of a complex, irregular shape, $\mathcal{L}$ can be taken of the order of the smallest semiaxis (several Mpc).
On the other hand, by using the same volume for both consecutive iterations, we ensure that we are detecting the mass elements that are being dynamically accreted onto the void, and not accounting for the elements which may appear inside or outside the ellipsoid due to its change between iterations.

We have also taken a conservative approach for the sake of showing the robustness of our results. Since DM particles can be traced, we can compare the total mass accretion in a certain redshift interval, when computed using our pseudo-Lagrangian method, and when computed by tracing the actual evolution of the particles in the simulation, i.e., checking which particles were outside the volume at the first iteration of the interval, and are inside that same volume at the last iteration. The results, shown in Supplementary Figure \ref{fig:supplementary_pseudolagrangian}, confirm that: (i) The pseudo-Lagrangian approach works well, on a statistical level, on the DM particle distribution, with reasonable scatter (0.3-0.5 dex) between the estimated and the actual accretion inflow; and (ii) On these low-density environments, gas and DM dynamics exhibit remarkably similar results, producing a scatter between DM and gas accretion rates typically below 0.2 dex. This serves as a confirmation of the applicability of our pseudo-Lagrangian approach for estimating gas accretion.

%%%%%%%%%%%%%
\section{Results} 
\label{sec:results}
%%%%%%%%%%%%%

\begin{figure}
\centering
{\includegraphics[width=0.4\textwidth]{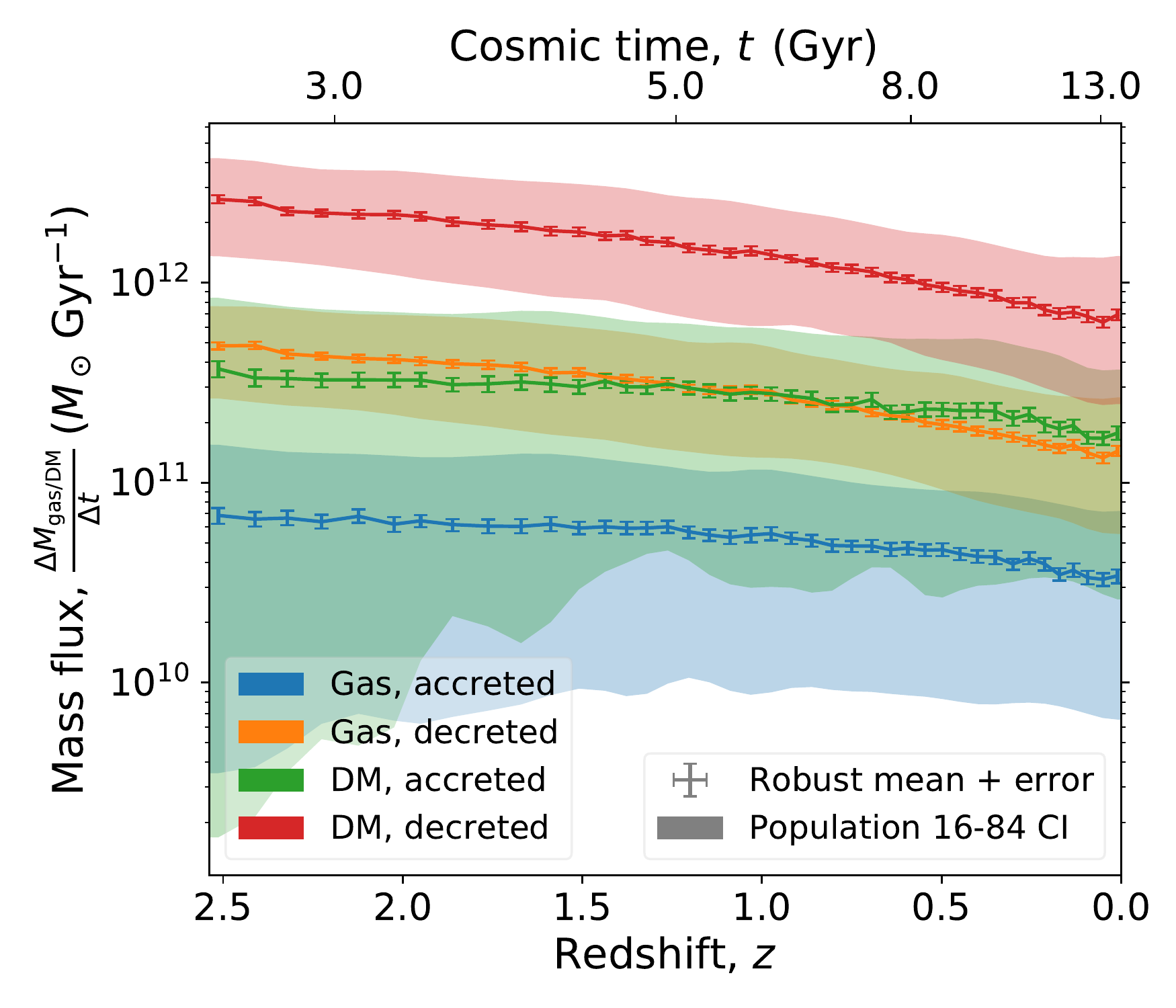} \label{fig:massfluxes.a}}\\
{\includegraphics[width=0.4\textwidth]{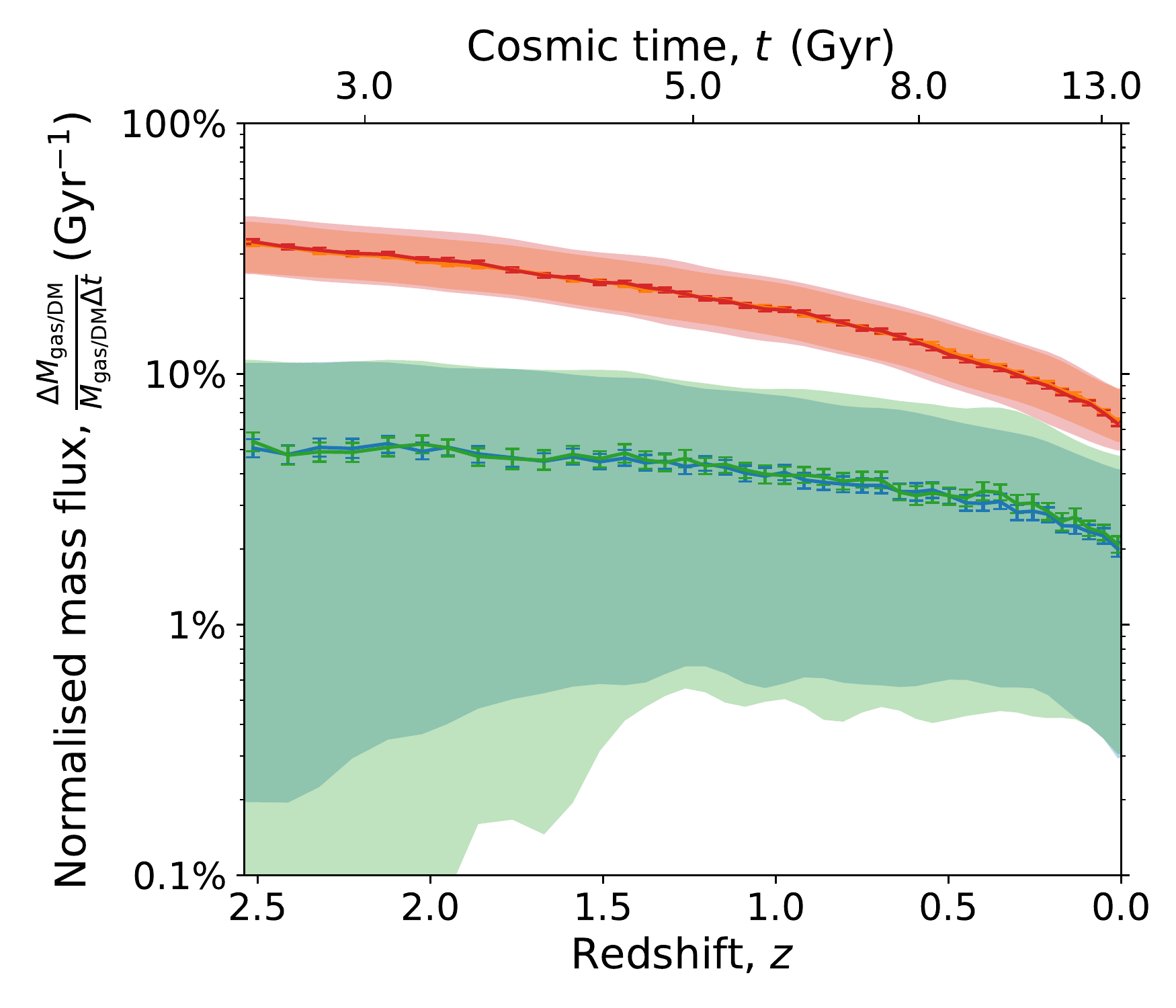} \label{fig:massfluxes.b}}\\
{\includegraphics[width=0.4\textwidth]{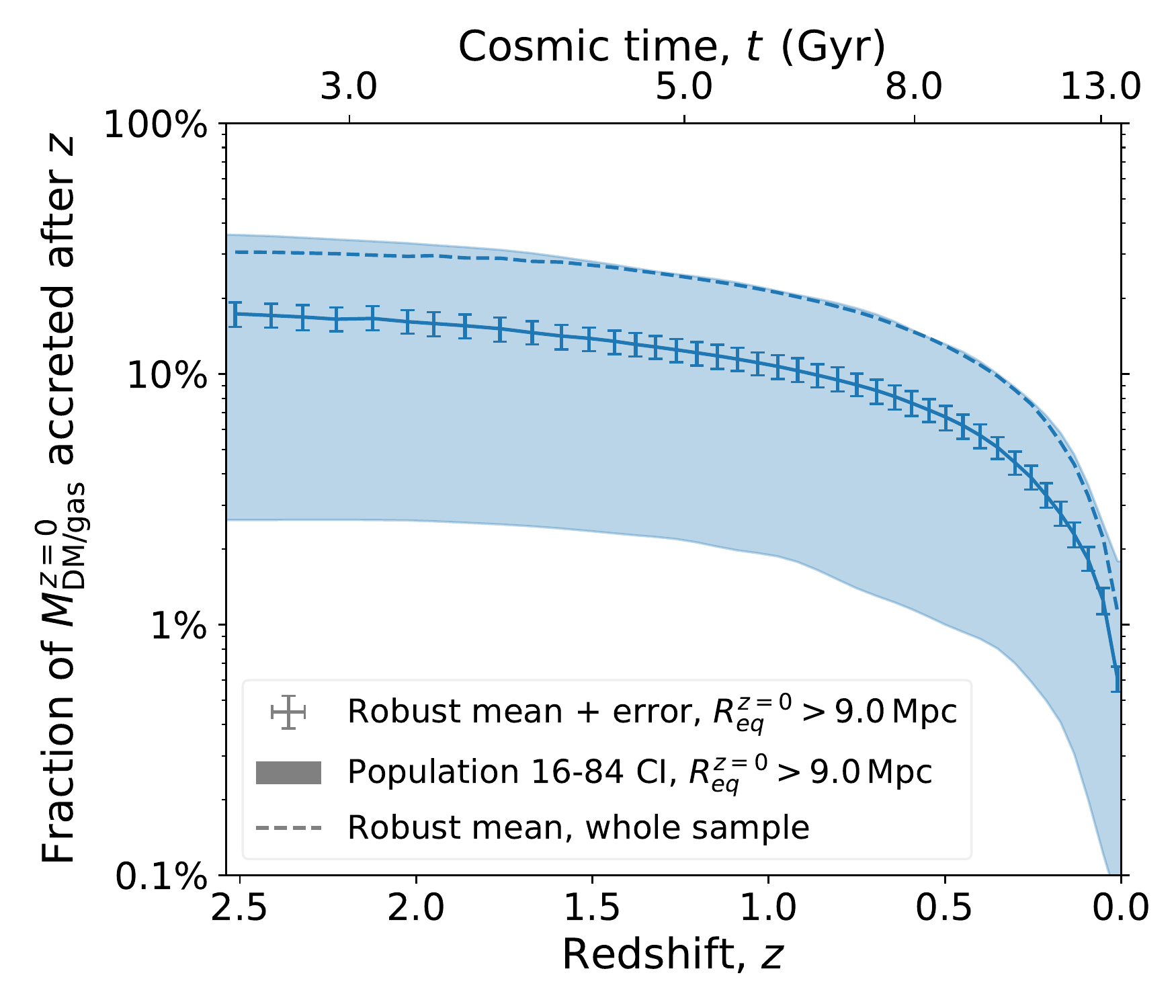} \label{fig:massfluxes.c}} 
\caption{Evolution of the accretion and decretion rates in the void sample. \textbf{Top panel:} evolution of the mass fluxes (gas and DM mass, entering or leaving the void per unit time, according to the colour legend) as a function of cosmic time. The shaded regions delineate the 16-84 percentiles (referred to as `confidence intervals' [CI] in the legend), as an indication of the scatter in these quantities, while the error bars represent the error of the mean value. \textbf{Middle panel:} mass fluxes normalised to the voids gas or DM mass, averaged over the cluster sample. Same legend as above. \textbf{Bottom panel:} anti-cummulative accreted gas mass fraction, i.e., the mass that has been accreted after a given redshift $z$ as a fraction of the gas mass of the void at $z=0$. Errors are given only for the $R_\mathrm{eq}^{z=0} > 9 \, \mathrm{Mpc}$ for clarity, being similar in magnitude for the whole sample.}
\label{fig:massfluxes}
\end{figure}

A summary of the results of this analysis over the whole cluster sample is shown in Fig. \ref{fig:massfluxes}, where we also present the decretion rates for comparison. The raw fluxes (gas or DM mass, entering or leaving the void per unit time) are displayed in the top panel, where it can be seen that accretion\footnote{Although these mass inflows would not be triggered by the peculiar gravitational fields, but by the external, large-scale structure bulk and shear velocity flows, since they move inwards in the voids and remain within them for long times, we refer to them as accretion flows in analogy with mass flows in massive objects.} flows onto voids, while smaller in magnitude than the decretion flows typically by a factor $1/6-1/3$, are present in the void sample in a statistical sense. 

In the middle panel we have normalised these fluxes to the mass of the given material component (gas or DM) in the void at each time, to be read as the percentage of gas or DM void's mass that leaves or enters the void per gigayear. By performing this normalisation, the robust mean values of the fluxes of DM and gas match each other, reflecting that both components undergo remarkably similar dynamics. This is expected, since gas in these regions has low temperature and pressure and thus behaves closer to a collisionless fluid (further validating the applicability of the pseudo-Lagrangian approach; see Sec. \ref{s:methods.pseudolagrangian}). While normalised decretion flows show little scatter, and evolve from nearly $30\% \, \mathrm{Gyr^{-1}}$ at $z=2.5$ to $\sim 6\% \, \mathrm{Gyr^{-1}}$ at $z\simeq 0$, accretion fluxes are smaller by a factor of $\sim 1/6$ at high redshift, but do not decrease as sharply and reach $\sim 1/3$ of the mean decretion values at $z=0$. It might be argued that part of these accretion flows could be due to void-in-cloud processes \citep[i.e., voids collapsing in a larger-scale overdense environment]{Sheth_2004, Sutter_2014}, even though our sample building strategy, from $z=0$ backwards in time, should exclude most of them since they would not have survived until $z=0$. Nevertheless, we have checked that the same results hold when restricting the sample to large voids ($R_\mathrm{eq}^{z=0} > 9 \, \mathrm{Mpc}$) only with a slight decrease in the accretion rates, less than a factor of 2, at low redshifts with respect to the whole sample. At high redshifts, there are no differences between large voids and the whole sample. Therefore, the accretion signal detected here cannot be ascribed to cloud-in-void processes alone. Smaller voids may experience stronger inflows, since they are more sensitive to external influences by a larger-scale velocity field.

Last, the bottom panel presents the accumulated accreted mass, as a function of the present-day voids gas or DM mass, from a redshift $z$ up to $z=0$. When focusing on the large-voids subsample, on average, up to $17\%$ of the voids current mass has been accreted after $z=2.5$ (reaching beyond $35\%$ at percentile 84), and the average void has suffered a mass inflow $10\%$ of its current mass after $z=1$. Interestingly, in their general analysis of the cosmic web, \cite{Cautun_2014} found that $\sim 20\%$ of the mass in voids at $z=0$ belonged to walls and filaments at $z=2$. Despite the similarity of the result, note that their interpretation is subtly different: while \cite{Cautun_2014} ascribe this result to an artefact due to the difficulty of identifying tenuous structures within voids as they become emptier, our result corresponds to an actual inflow (matter initially outside the void, which crosses its boundary at a given time).

\begin{figure*}
\centering
\includegraphics[width=0.6\textwidth]{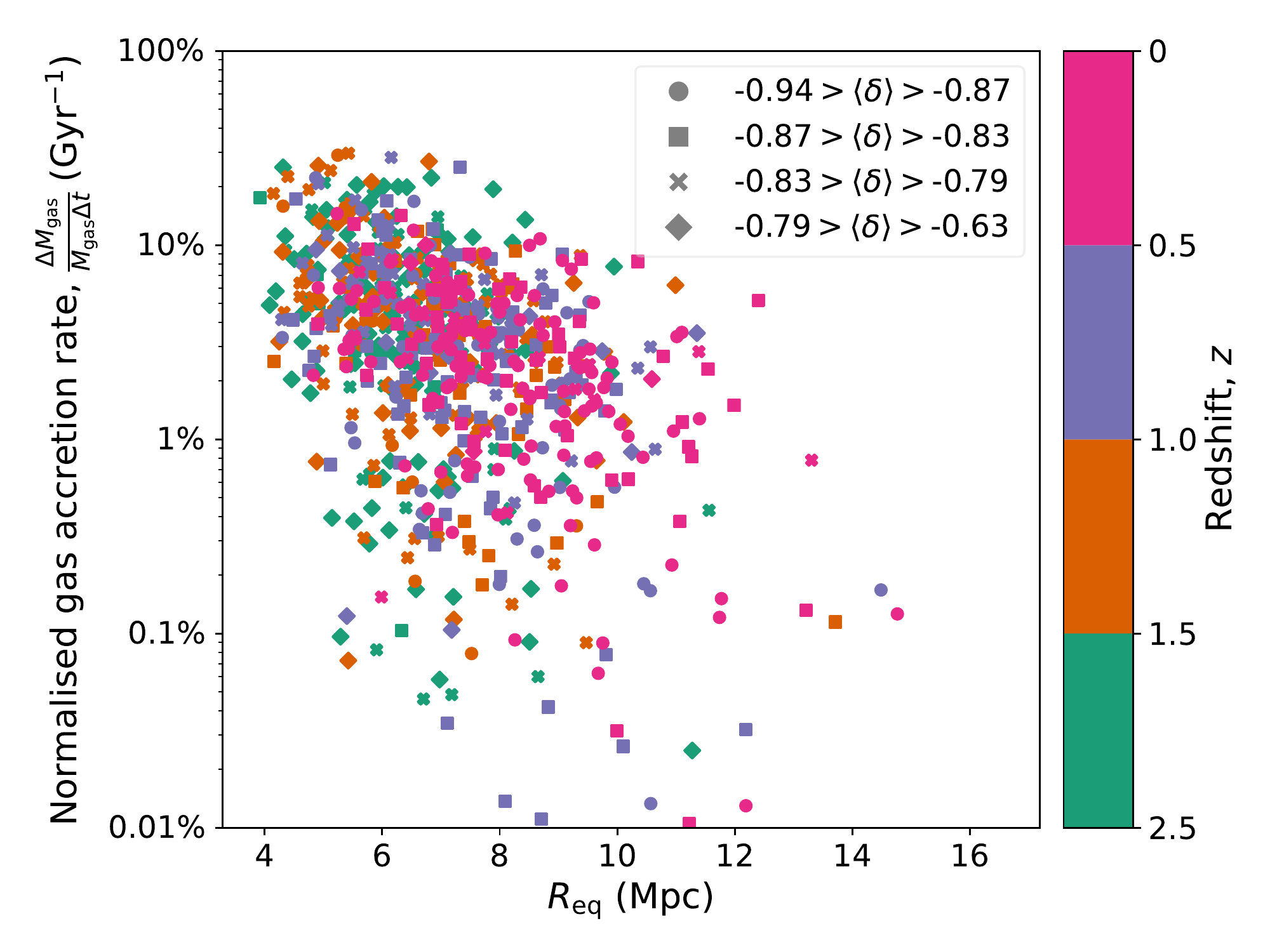}
\caption{Normalised gas accretion rate, i.e., gas inflow per Gyr in units of the void's gas mass (vertical axis) as a function of the size of the void (equivalent radius, horizontal axis). The gas accretion rates are computed on four redshift intervals, encoded in the figure according to the colour scale on the right. The shapes of the data points refer to the mean overdensity of the void, according to the binning specified in the legend.}
\label{fig:scatter}
\end{figure*}

As the central result of this Letter, in Fig. \ref{fig:scatter} we present the gas accretion rates (gas mass accreted, normalised by the voids' mass and per unit time) as a function of the void size (equivalent radius), computed on four redshift intervals (from $z=2.5$ to $z=1.5$, and three subsequent intervals with $\Delta z = 0.5$ thereon) which are encoded by the colourscale in the figure. A significant fraction of voids, at any redshift, presents relevant accretion rates (above a few percents per gigayear, which are sufficient to impact their composition and dynamics). The highest accretion rates are seen in the smallest voids. As mentioned above, smaller voids are more prone to externally-induced flows due to larger-scale influences: while $R_\mathrm{eq} \sim 5 \, \mathrm{Mpc}$ voids show mean values $\sim 8 \% \, \mathrm{Gyr^{-1}}$, this rate lowers to $\sim (1-2) \% \, \mathrm{Gyr^{-1}}$ in the case of the largest voids. Nevertheless, large voids with exceptionally large accretion rates also exist, even at low redshifts. Naturally, the abundance of small voids makes it possible for some of them to show extreme accretion rate values. The point markers encode the mean overdensity, $\langle\delta\rangle \equiv \langle\rho\rangle / \rho_B - 1$ of the voids, according to the legend, to check whether there is any trend between this property and the accretion rates. All the voids in the sample, and even the large and rapidly-accreting ones, have very small overdensities, thus ruling out the fact that our results could be contaminated by a bad delineation of the voids wall. Indeed, as discussed in Sec. \ref{s:methods.void finder}, our void identification technique has aimed to be conservative enough to exclude these possible effects. 

\begin{figure*}
\centering
\includegraphics[width=0.7\textwidth]{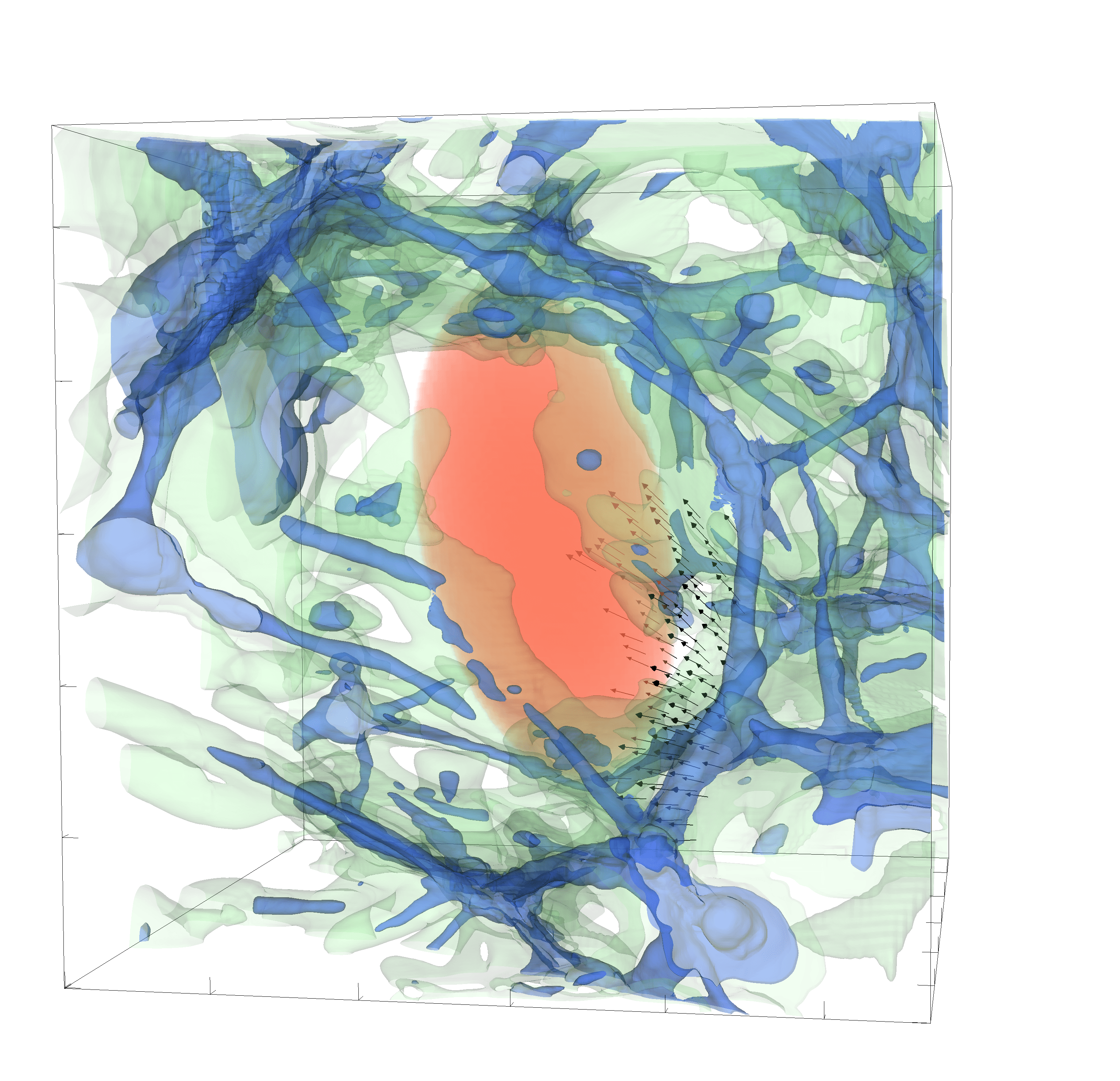}
\caption{Close look at a large ($R_\mathrm{eq} \simeq 12 \, \mathrm{Mpc}$) void which undergoes significant accretion by redshift $z \simeq 0$. The box corresponds to a cubic domain $\sim 56 \, \mathrm{Mpc}$ (comoving) along each direction. The ticks on the axes are spaced 10 Mpc for visual reference. Blue and pale green contours represent the gas density, with blue and green regions approximately corresponding to collapsed (cluster and filaments) matter and diffuse gas around it, respectively. The orange, shaded area shows the location of the ellipsoidal void, which is well delineated by the surrounding matter. The arrows represent the gas velocity field around the surface of the void in those regions where gas is being accreted towards it.}
\label{fig:1void}
\end{figure*}

To better visualize the effect, we show in Fig. \ref{fig:1void} the gas density field around a large void ($R_\mathrm{eq} \simeq 12 \, \mathrm{Mpc}$) with high accretion rates. Blue and green contours correspond to iso-density surfaces of $\rho / \rho_\mathrm{B} \simeq 3 \; \text{and} \; 0.5$, roughly enclosing mean total densities $15\rho_B$ and $4 \rho_B$, respectively, in order to give context of the distribution of matter around the void. The orange shadow highlights the void, with mean total density $\rho/\rho_B \simeq 0.1$. On top, we overplot the accretion velocity field, that is, the velocity vectors in the region around the void where they point towards it. This representation clearly exemplifies the presence of coherent, large-scale streams of matter flowing towards cosmic voids from higher-density regions. Complementarily, in Supplementary Figure \ref{fig:supplementary_movie} we show a movie of a gas density slice through a large ($R_\mathrm{eq} \simeq 14 \, \mathrm{Mpc}$) void which undergoes significant accretion, with the velocity field overplotted with arrows. The movie exemplifies the nature of these inflows: the velocity field around the void consists on its own induced outflow (dominant through most of its boundary), plus the bulk and shear flows originated by the surrounding structures (see also \citealp{Aragon-Calvo_2013} for a detailed study of the hierarchical nature of velocity fields in and around voids).

Tracing the newly accreted DM particles in time, we find that nearly $50\%$ remain inside the original volume of the void for up to $\sim 10 \, \mathrm{Gyr}$, and a significant fraction of them reach inner radii. The same behaviour is expected for the gas, therefore granting the accreted gas, which may have been pre-processed outside the void, a long enough timespan to play a crucial role in the formation and evolution of void galaxies. 

%%%%%%%%%%%%%
\section{Conclusions} 
\label{sec:conclusion}
%%%%%%%%%%%%%

The findings reported in this Letter challenge the common accepted picture on the evolution of cosmic voids and could consequently have a direct potential impact on the understanding of galaxy formation and evolution in low-density environments. Hence, the uncontaminated and pristine void domains could be altered by the entrance of chemically and thermodynamically processed gas. Future effort should be devoted to confirm these results with other simulation codes and void identification strategies (e.g., those based on the watershed transform \citealp{Platen_2007, Neyrinck_2008}, or other, see for example the comparison project of \citealp{Colberg_2008}).

%%%%%%%%%%%%% ACKNOWLODGEMENTS
%%%%%%%%%%%%%%%%%%%%%%%%%%%
\subsection*{Acknowledgements}
%%%%%%%%%%%%%%%%%%%%%%%%%%%
The authors thank the referee for his/her constructive criticism. This work has been supported by the Spanish Agencia Estatal de Investigaci\'on (AEI, grant PID2019-107427GB-C33) and by the Generalitat Valenciana (grant PROMETEO/2019/071). DV acknowledges partial support from Universitat de Val\`encia through an \textit{Atracció de Talent} fellowship. Simulations have been carried out using the supercomputer Llu\'is Vives at the Servei d'Inform\`atica of the Universitat de Val\`encia.
%%%%%%%%%%%%%

%\bibliography{apjl_voids}{}
 \newcommand{\noop}[1]{}

\bibliographystyle{aasjournal}

%%%%%%%%%%%%%%%%%%%%%%%%%%%%%%%%%%%%%%%%%%%%%%%%%%%%%%%%%%%%%% SUPPL MATERIAL
\appendix
\section{Supplementary material}
\renewcommand{\figurename}{Supplementary Figure} 
\renewcommand{\tablename}{Supplementary Table}
\setcounter{figure}{0}
\begin{figure*}[!ht]
\centering
\includegraphics[width=.85\textwidth]{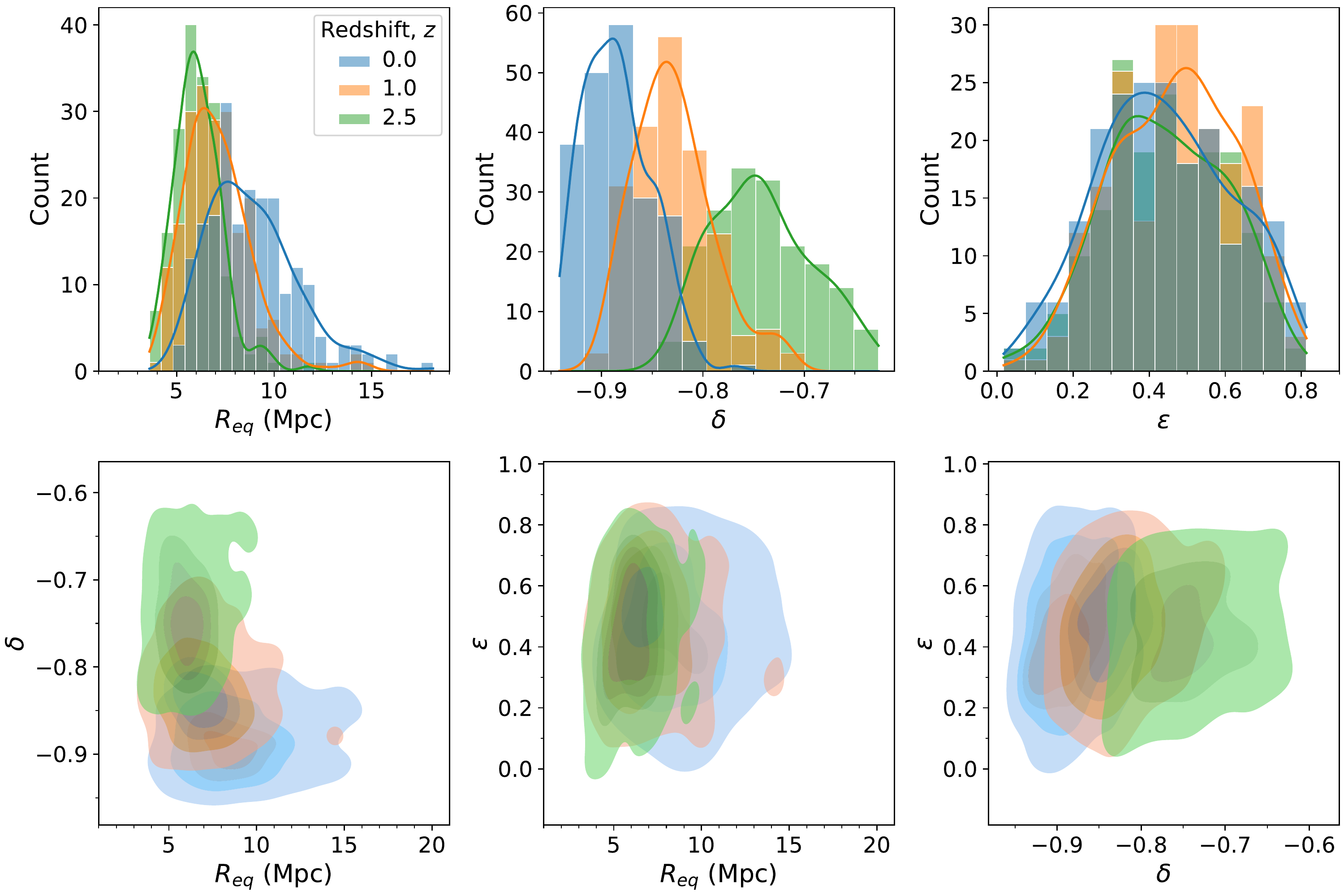}
\caption{Summary of general properties of the void sample. \textbf{Top row:} distribution of the equivalent radius ($R_\mathrm{eq}$, left panel), mean overdensity ($\delta$, middle panel) and ellipticity ($\epsilon$, right panel) for the voids sample at redshifts $z\simeq 0$ (blue), $1$ (orange) and $2.5$ (green). \textbf{Bottom row:} joint distribution of each pair of variables, according to the same color pallete. The continuous lines in the histograms, and the color contours in the joint distributions have been obtained by means of a Gaussian kernel density estimation procedure. Darker colours imply higher density of voids in the corresponding parametric space.}
\label{fig:supplementary_sample}
\end{figure*}

\begin{figure*}[!ht]
\centering
\includegraphics[width=0.4\textwidth]{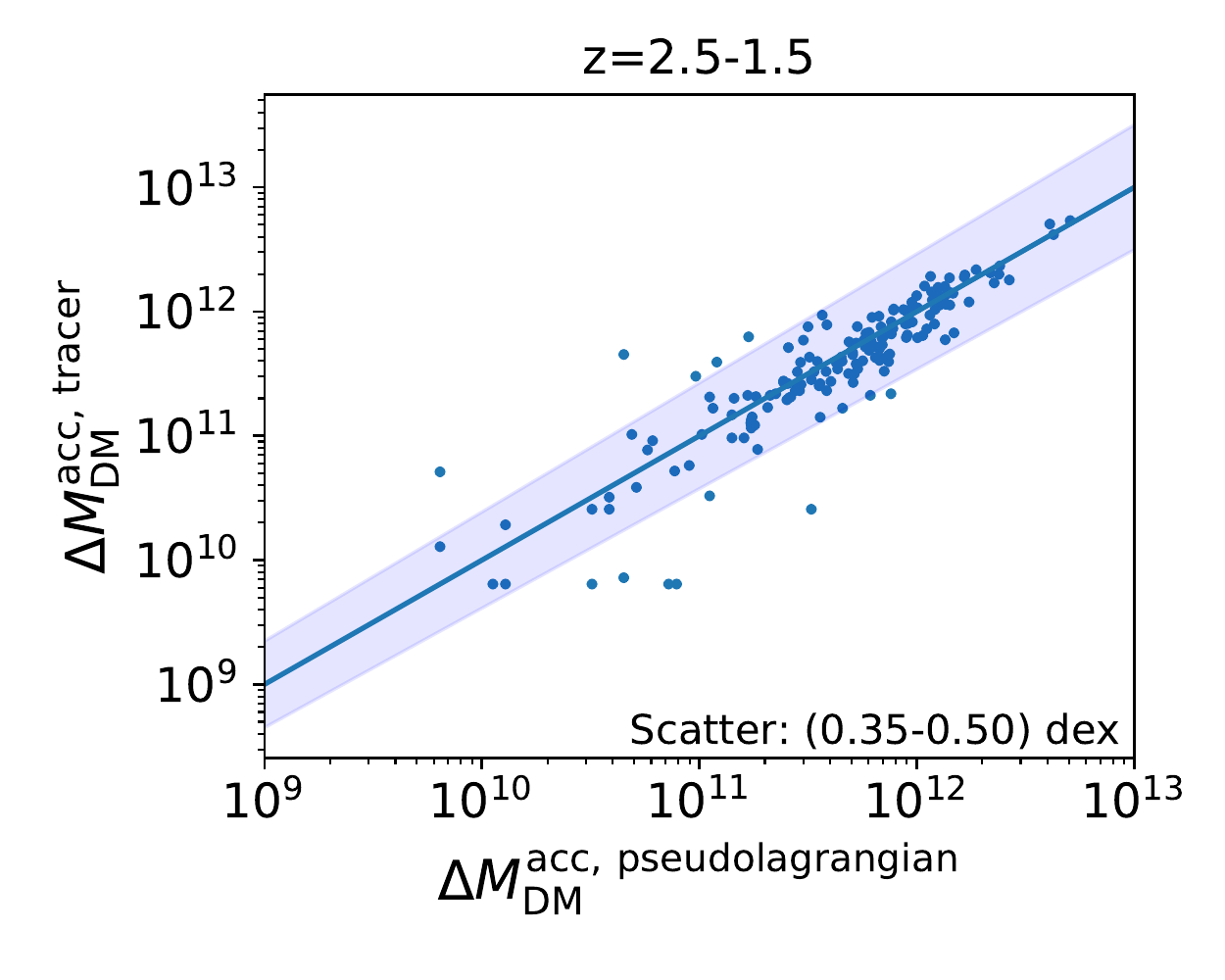}~ 
\includegraphics[width=0.4\textwidth]{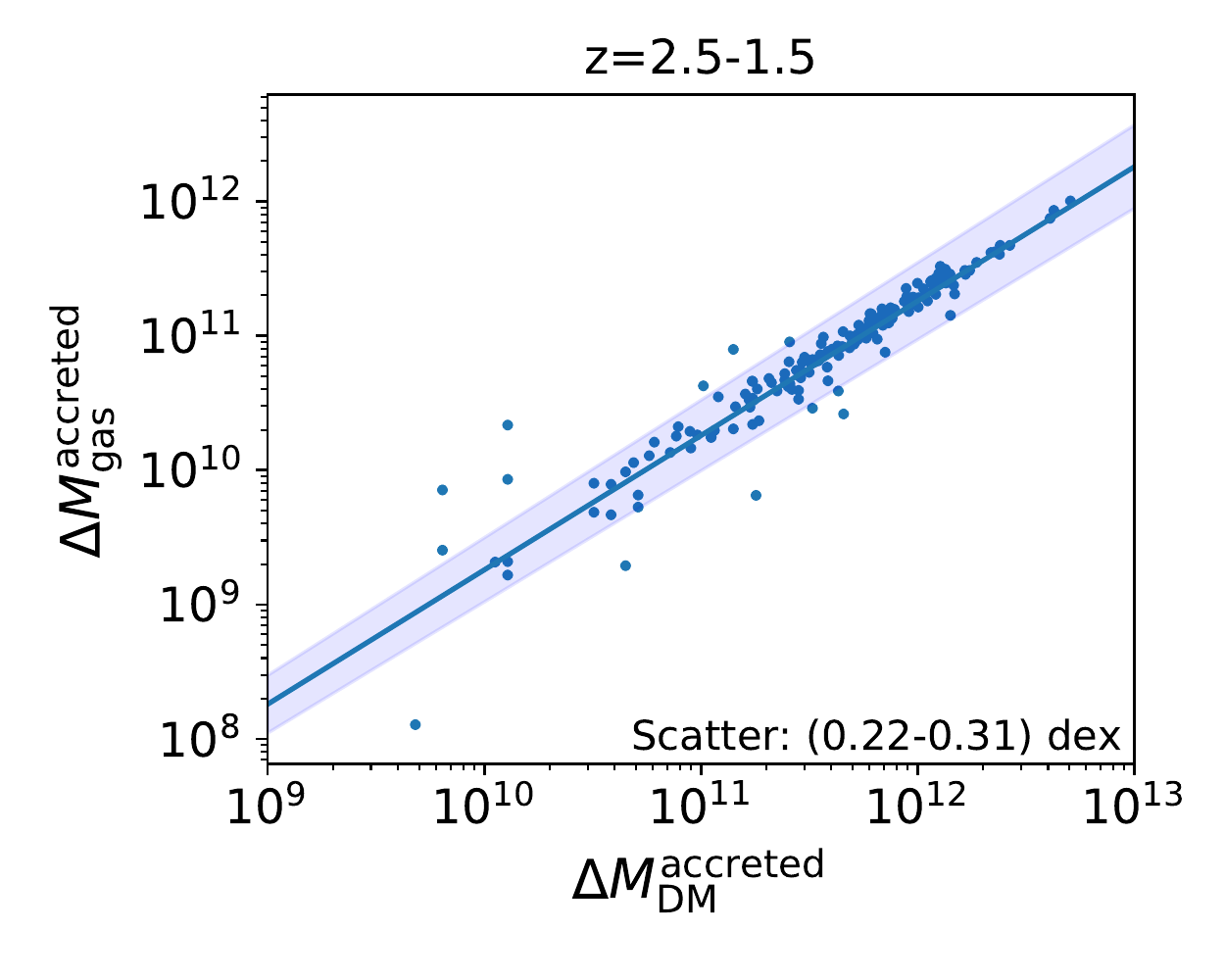}
\includegraphics[width=0.4\textwidth]{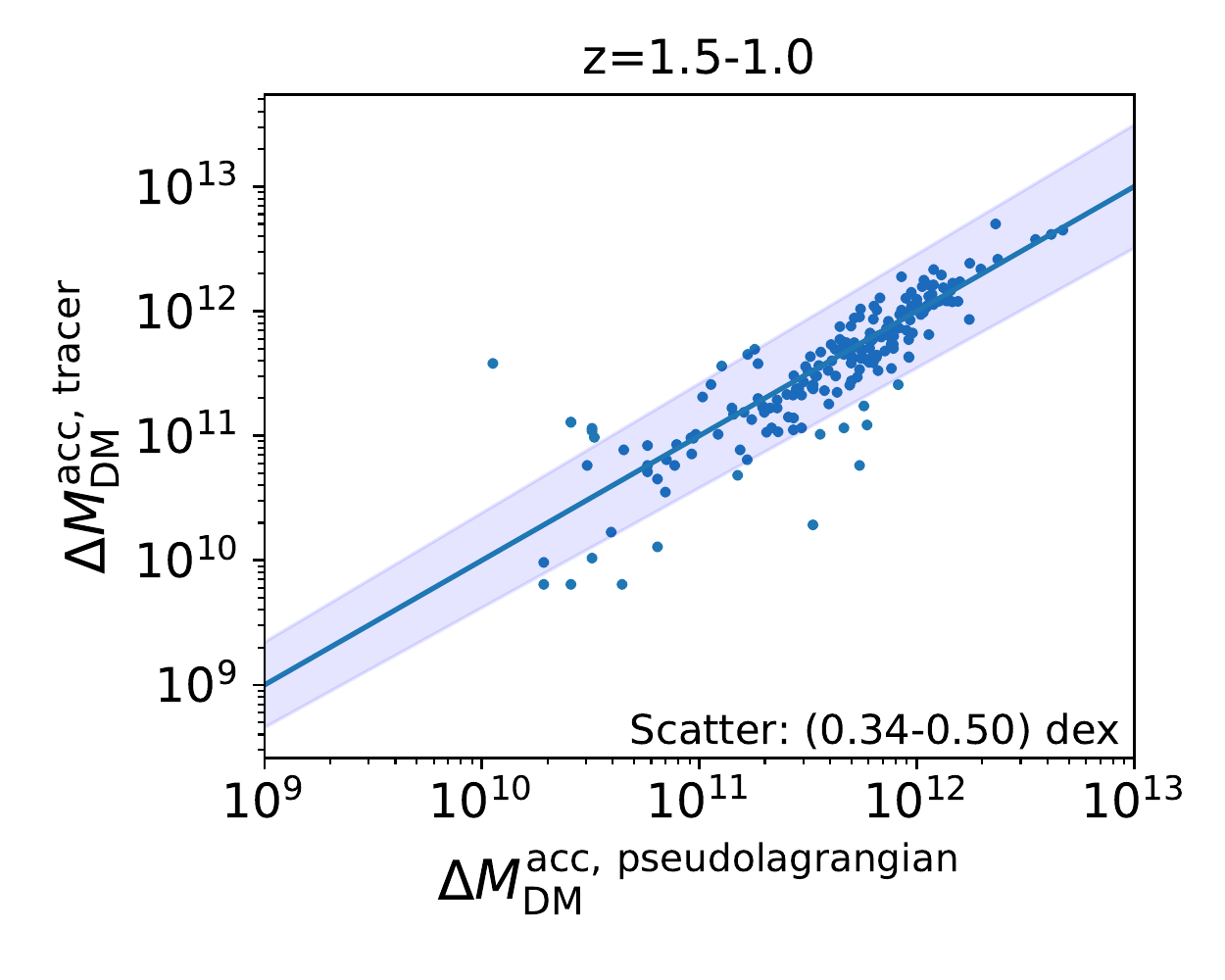}~ 
\includegraphics[width=0.4\textwidth]{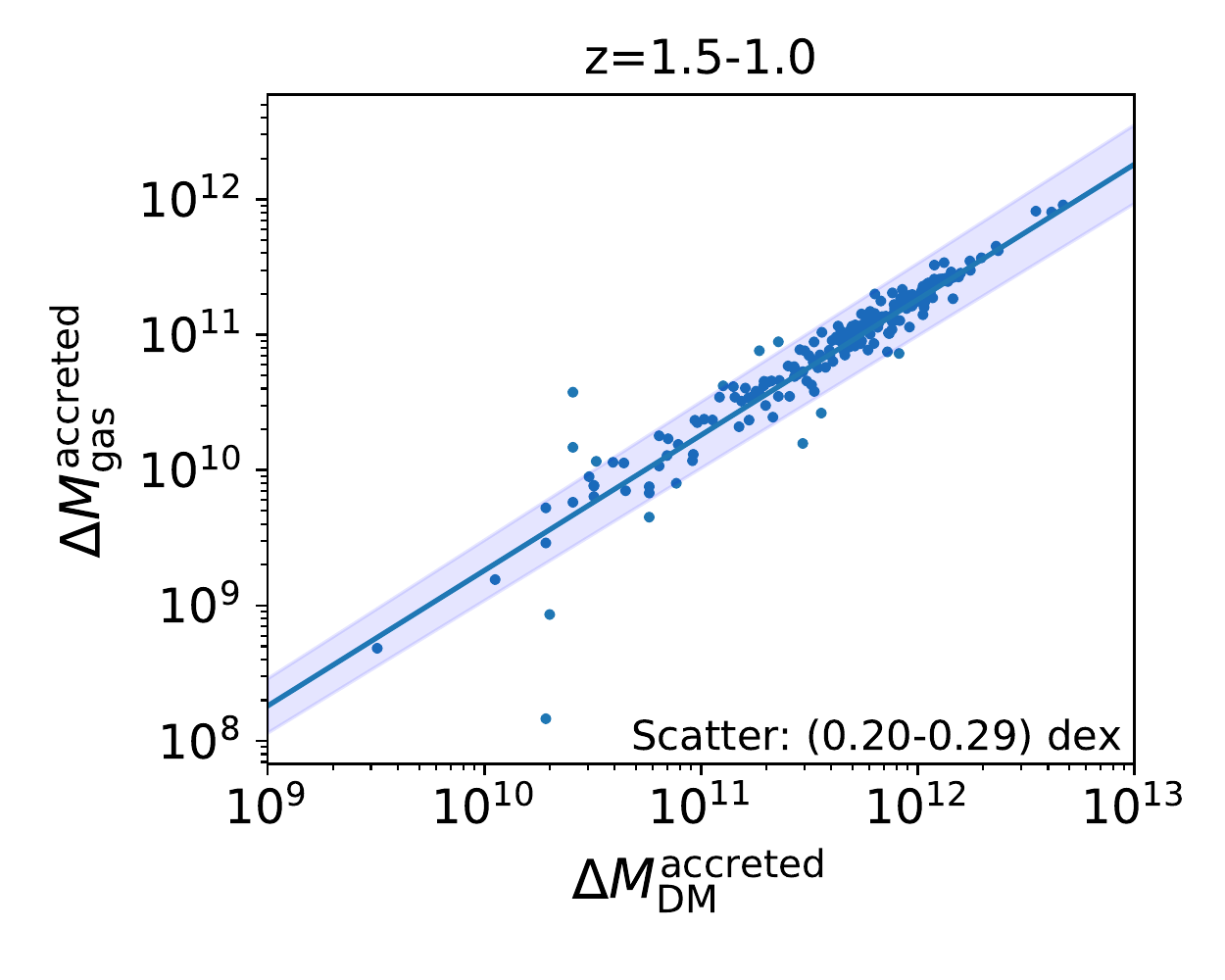}
\includegraphics[width=0.4\textwidth]{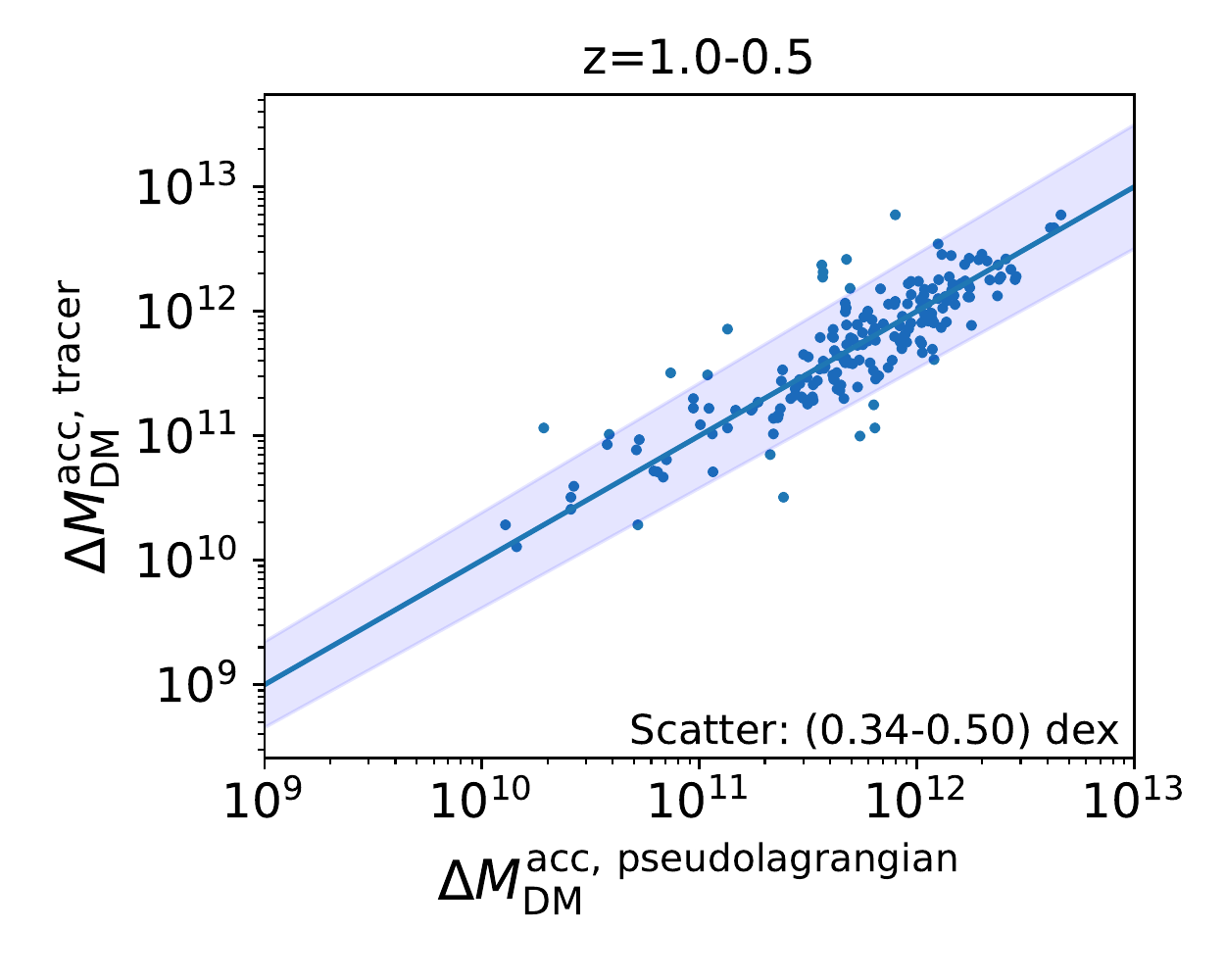}~ 
\includegraphics[width=0.4\textwidth]{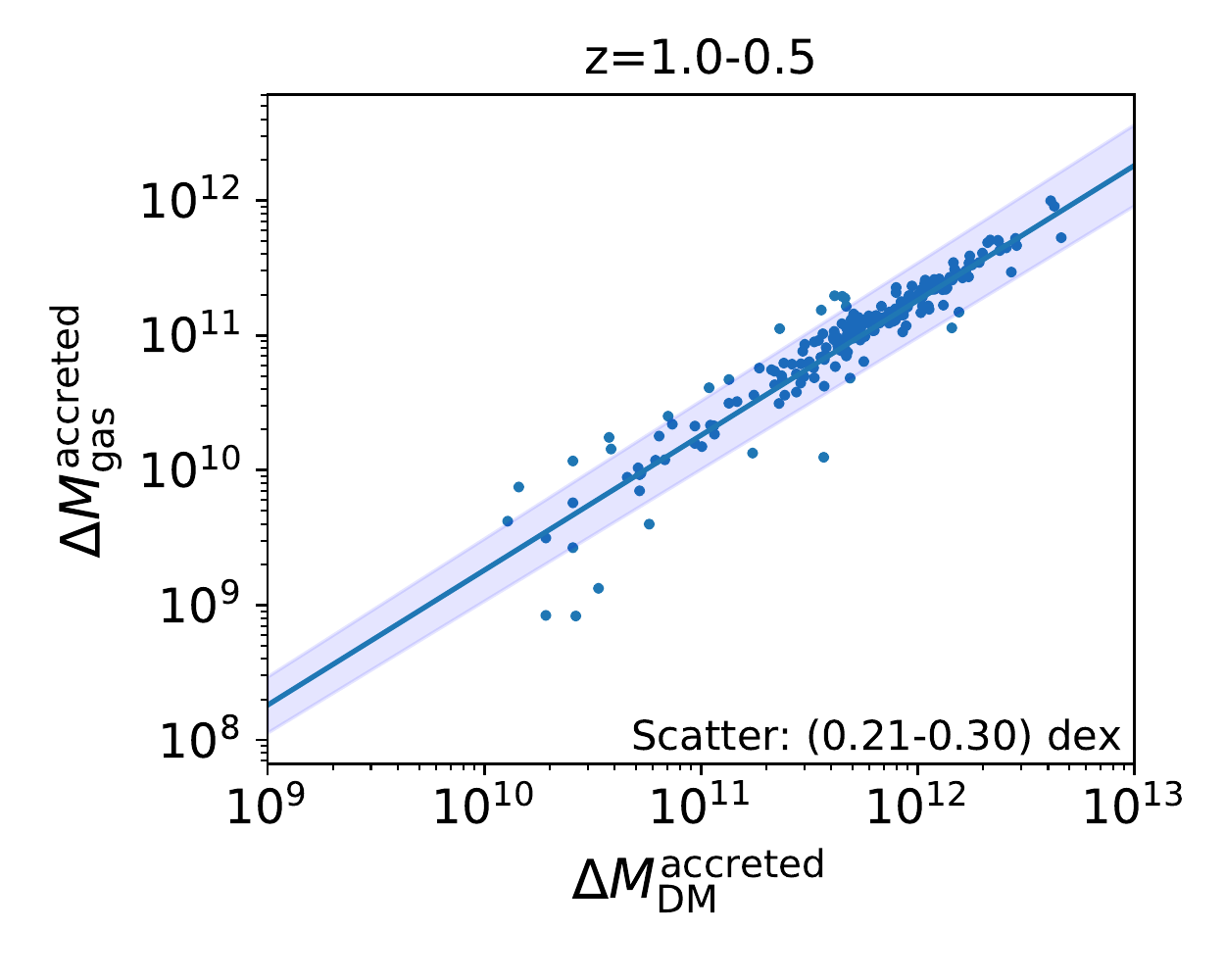}
\includegraphics[width=0.4\textwidth]{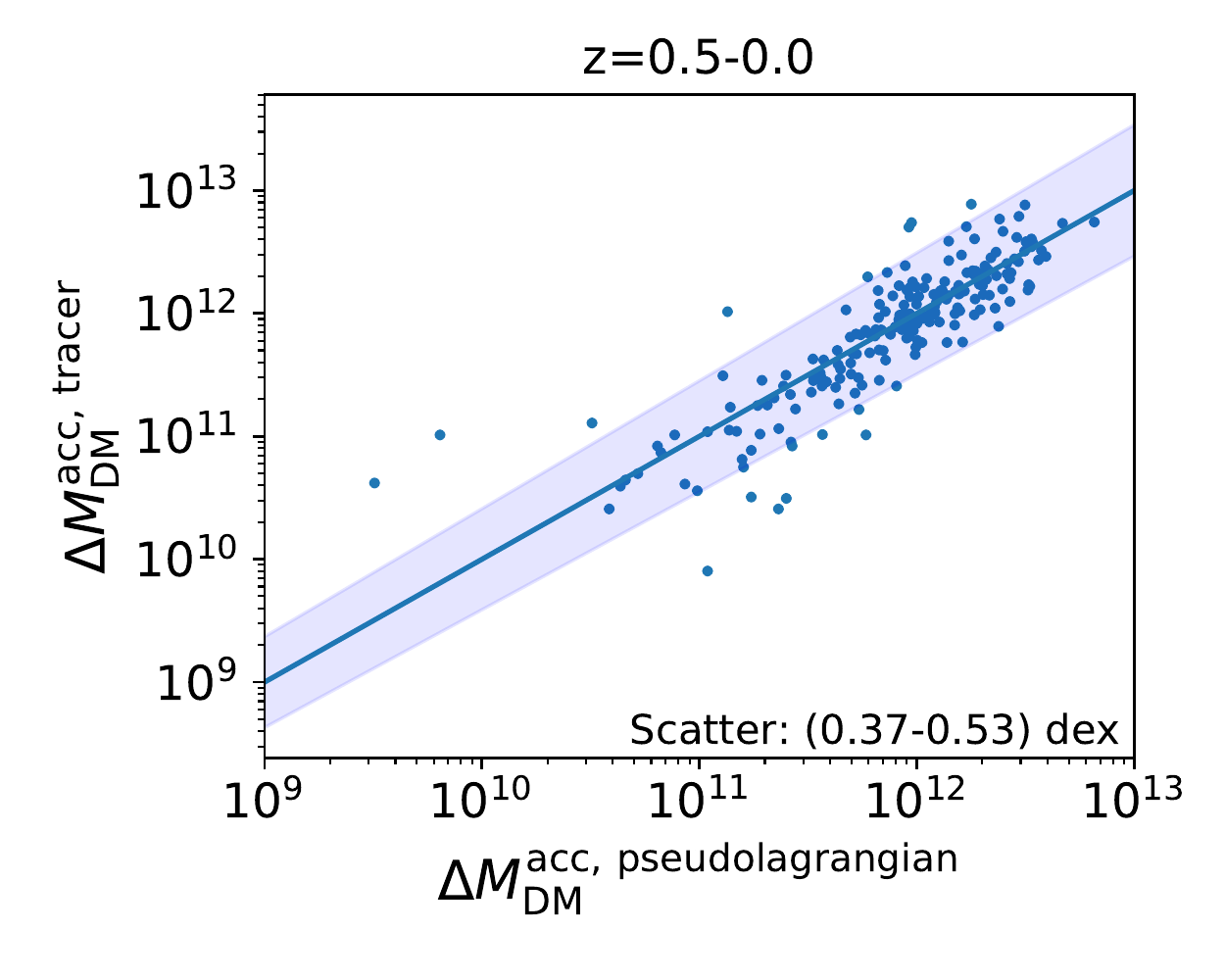}~
\includegraphics[width=0.4\textwidth]{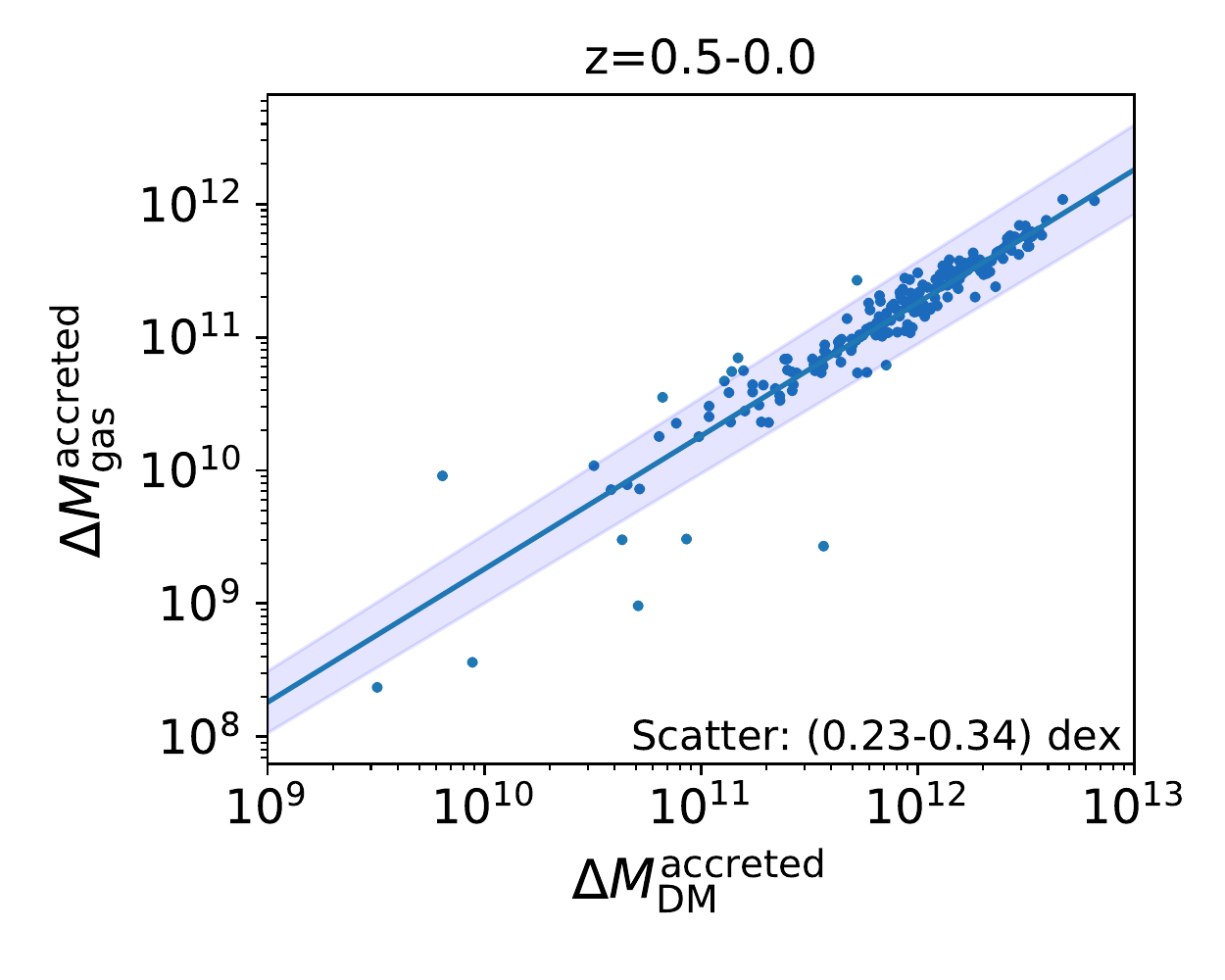}
\caption{\textbf{Left-hand side column:} correlation between the accreted DM mass computed according to our pseudo-Lagrangian method, and when computed by explicitly tracing DM particles in the simulation, in the redshift interval specified at the top of each panel. \textbf{Right-hand side column:} tight correlation between the gas and the DM accreted mass, both computed according to our pseudo-Lagrangian algorithm.}
\label{fig:supplementary_pseudolagrangian}
\end{figure*}

\clearpage

\begin{figure}
\begin{interactive}{animation}{figS3_animation}
\includegraphics[width=0.75\textwidth]{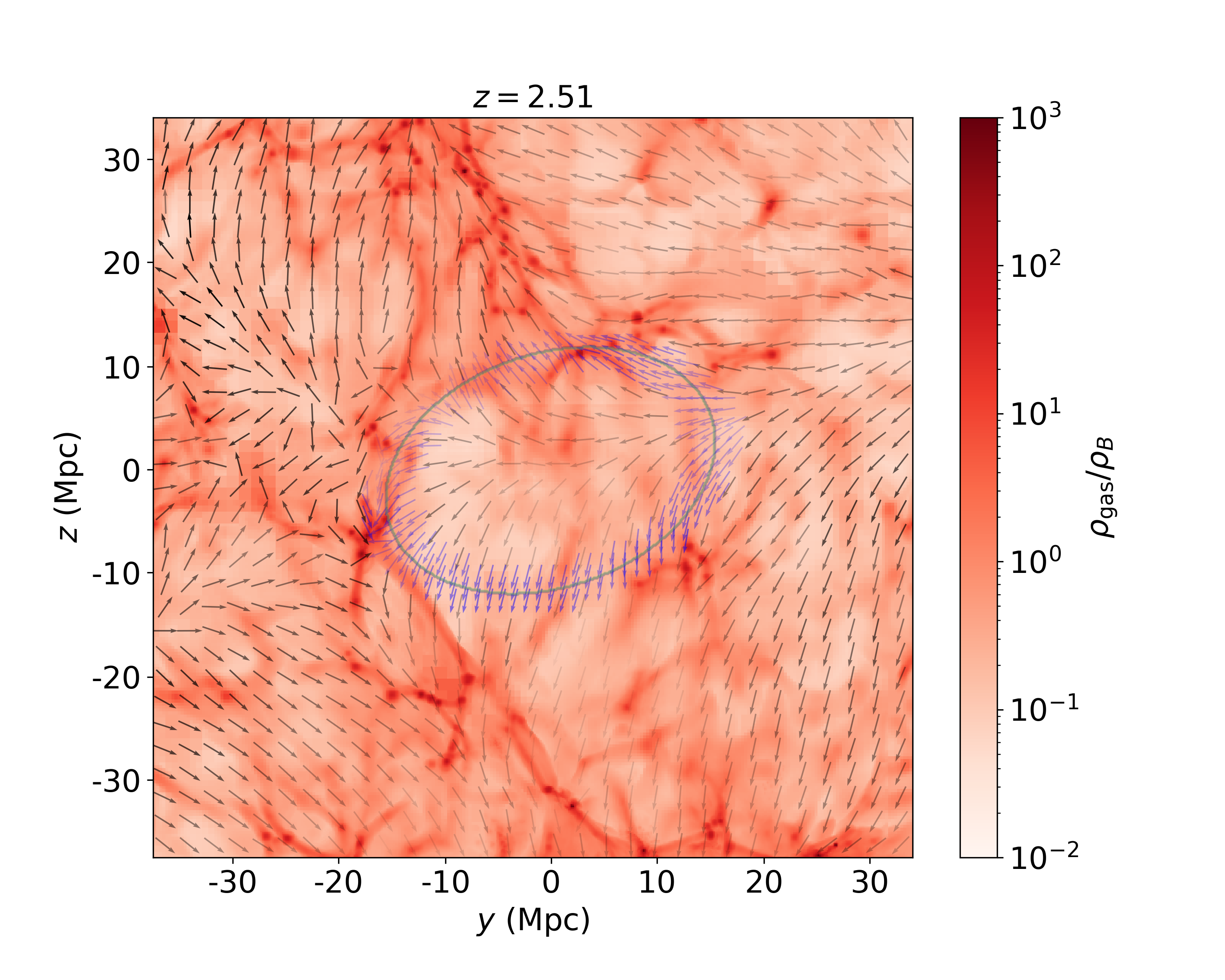}
\end{interactive}
\caption{Movie showing the evolution of a density slice ($\sim 9 \, \mathrm{Mpc}$ thick) through a large ($R_\mathrm{eq} \simeq 14 \, \mathrm{Mpc}$) void and its environment, from $z=2.5$ to $z=0$. The colours encode the gas density (in units of the background density) according to the adjacent colour scale. The green ellipse represents the slice through the void's ellipsoid. Note that some part of the void's wall may appear to be inside the void due to projection effects. Nevertheless, the void overdensity is low, below $\langle \delta \rangle \sim -0.8$, thus ensuring that overdense regions are well excluded. Arrows represent the gas velocity field in the slice. Note that the size of the arrows is constant for better visualization, while the magnitude of the projected velocity vector is encoded in the opacity of the arrow (more opaque arrows imply higher velocity magnitudes). The velocity field around the void's boundary is sampled with higher density of arrows and coloured blue for better visualization. Matter entering the void through its rightmost corner can be clearly visualized in the movie. This figure is available as an animation in the HTML version of the article. \textit{Still frame}: first frame of the movie. \textit{Duration}: 8 seconds.}
\label{fig:supplementary_movie}
\end{figure}

\end{document}